# 2.5 eV Pulsed Cathodoluminesce band of silicon dioxide

V.A. Kozlov [a], S.A. Kutovoi [b], N.V. Pestovskii [a),c)], A.A. Petrov [a),c)], A.A. Rodionov [a),c)],

S.Yu. Savinov [a),c)], Yu.D. Zavartsev [b], M.V. Zavertyaev [a] and A.I. Zagumennyi [b]

[a] P.N. Lebedev Physical Institute of the Russian Academy of Sciences, 119991, 53 Leninskiy Prospekt, Moscow, Russian Federation.

[b] A.M. Prokhorov Institute of General Physics of the Russian Academy of Sciences, 119991, 38 Vavilov Str., Moscow, Russian Federation.

[c] Moscow Institute of Physics and Technology (State University), 141700, 9 Institutskiy per., Dolgoprudny, Moscow Region, Russian Federation.

**Abstract**

Room-temperature (RT) Pulsed Cathodoluminescence (PCL) spectra of a set of pure synthetic (both crystalline and amorphous) silicon dioxide materials were studied. It is shown, that the PCL spectra of all samples (both amorphous and crystalline) possess a separate band at 495 nm (2.5 eV). This band is the most intensive one in PCL spectra of disordered materials. The RT PCL band at 495 nm (2.5 eV) of α-quartz single crystal is polarized in XY crystalline plane (perpendicular to the 3$^{rd}$ order symmetry axis). The structure of this band was detected. It consists of three peaks: at 480±2 nm (2.58±0.01 eV), 487±2 nm (2.55±0.01 eV) and 493±2 nm (2.52±0.01 eV). Energy separation between peaks is equal in order of magnitude to energies of $Li_xO_y$ molecular vibrations and to the energy of optical phonon in α-quartz. It is shown, that the emission band at 495 nm (2.5 eV) in RT PCL spectra of α-quartz single crystal is related to the bulk emission centers, not to the surface-related ones. The annealing behaviors of the 495 nm (2.5 eV) bands in spectrum of amorphous and crystalline $SiO_2$ are close to each other. This fact may be the manifestation of identical origin of these bands. The following explanation of experimental data is proposed: the origin of 495 nm (2.5 eV) band in pure silicon dioxide is related to the recombination of non-bridging oxygen $NBO^-$-$Li^+$ centers.

**Corresponding author**: Nikolai V. Pestovskii

**e-mail**: pestovsky@phystech.edu

**postal address**: 119991, 53 Leninskiy Prospekt, Moscow, Russian Federation.

**Highlights**:

1. The RT PCL band at 2.5 eV is studied in spectra of both amorphous and crystalline $SiO_2$.
2. The structure with three peaks and polarization in XY crystalline plane of 2.5 eV band in RT PCL spectrum of α-quartz single crystal is found.
3. The α-quartz 2.5 eV RT PCL emission band is related to the bulk emission centers.
4. The annealing behavior of 2.5 eV PCL band is close in the case of amorphous and crystalline $SiO_2$.
5. The model suggested is based on the relation of 2.5 eV emission of $SiO_2$ to the $NBO^-$-$Li^+$ complexes recombination.

1. **Introduction**

Silicon dioxide is a key material for a wide range of optical and electronic applications [1]. Also, the substantial portion (~12 wt%) of the Earth's surface consists of various forms of crystalline silicon dioxide [2]. For these reasons, the luminescence of $SiO_2$ under the various excitations was extensively studied [1,3,4 and references therein]. However, it cannot be said at present, that the problem of physics of luminescence in $SiO_2$ is solved finally.

The separate band at 2.5 eV in cathodoluminescence (CL) spectra of pure crystalline quartz under the action of electron pulses with energies of 10 keV and duration of ~2 ms at room temperature (RT) was observed in [5]. The origin of this band did not been discussed. In [6] the wide single emission band at 2.6 eV was observed, and its polarization predominantly parallel to Z (3[rd] order symmetry) crystalline axis was detected. In [7] the X-ray induced luminescence of quartz was studied. Two overlapping bands having different polarization were observed. The positions of bands were 2.74 eV and 2.48 eV. It was shown, that the 2.48 eV-band was completely polarized in XY (perpendicular to Z) crystal plane (in Cartesian coordinates). The 2.74 eV band was polarized mainly parallel to the Z crystal axis. Before the X-ray luminescence studies, the sample was irradiated with 1.7 MeV electron beam in order to eliminate the emission band at 3.26 eV (380 nm) [7]. Two unresolved bands at 2.8 and 2.5 eV in luminescence spectra of quartz were observed in [8] under the action of high-current (~100 A) short (10-20 ns) electron beam with electron energies of ~2 MeV. According to [8], the excitation density achieved the level of $10^{18}$ cm$^{-3}$. The polarization of emission bands was similar to [7]. At the same time, the saturation of the 2.5 eV emission intensity was found at certain electron beam fluence. However, no saturation was found in the 2.8 eV band intensity. Based on this result it was concluded, that the band at 2.8 eV have the intrinsic origin and was attributed to the Self-Trapped Exciton (STE) recombination. The band at 2.5 eV was assigned as the extrinsic and was attributed to the defects or impurities [8].

The results of [9] contradicts to the conclusion of [8]. In [9] the saturation of 2.5 eV band intensity had not been found. An another model was proposed in [9], concerning two types of STE in α-quartz [9-12]. According to this model, STE creation is related to Si-O bond breaking or weakening and non-bridging oxygen (NBO) formation, which relaxes and makes a bond with bonding oxygen on the other side of Z or (X,Y) channel. These two oxygen bond orientations determine the final STE orientation and the corresponding polarization of luminescence. Thus, the emission at 2.5 eV was assigned in [9-12] to STE recombination.

The model of STE as the Si-O bond rapture, when the hole is located on the non-bridging oxygen (NBO) atom and the electron is located on the Si atom, was confirmed via the ab-initio calculations in [13]. However, the results of [13] predicted only one type of STE with an excellent coincidence with experimental data on the peak position (2.8 eV) and polarization degree. No STE emission at 2.5 eV was predicted [13]. At the same time, in [14] two types of STE were predicted, but the emission energies of their recombination were underestimated (2.01 and 2.46 eV) in comparison with the experimental data.

The introduction of $Na^+$, $Li^+$, $H^+$ ions into the quartz via the high-temperature thermodiffusion ("sweeping") was used in [15] to study the role of alkali ions in quartz luminescence. It was shown [15], that the intensity of 485 nm (2.53 eV) emission band under the X-ray excitation at RT increases with an increase of Li concentration. It was also mentioned [15], that the intensity of 485 nm (2.53 eV) emission band is presumably increased with the growth of irradiation dose in $Li^+$-excess samples. Authors [15] suggested a possible correlation of the 485 nm (2.53 eV) emission band with extrinsic defects, whose concentration is modified by sweeping. According to [15], the recombination of an electron and hole trapped adjacent to an $[AlO_4/M^+]$ center may occur [16], when $M^+$ is the $Li^+$ ion.

It is well-known [17,18], that the electrons with energies higher than the 200 keV produces the significant beam-induced damage in silicon dioxide (primarily, via the knock-on mechanism) [17]. According to this statement, the destruction of sample may lead to the distortion of luminescence spectra. In addition, it is worth to note, that the luminescence measurements were carried out in [8] ~20 µs after the irradiation of samples. Thus, the spectra of the initial stage of luminescence in [8] were not studied.

Review of published results shows, that the nature of 2.5 eV emission of silicon dioxide at present time is not clear. Moreover, the conclusions of previous works are contradictory.

In present work the pulsed cathodoluminescence (PCL) spectra of pure synthetic amorphous silica powder before and after the annealing, ultra-pure silicon dioxide powder after annealing, α-quartz single crystal before and after annealing and the perfect crystalline α-quartz powder were measured in the region of 200-800 nm. Energies of particles in incident beam were of ~150 keV. Firstly, the comparison of PCL spectra of all materials will be presented in the region of 200-800 nm. Secondly, the detailed study of α-quartz single crystal PCL spectrum will be performed. All spectra were measured at room temperature and at ambient-air.

The purpose of this study is to explore the properties of 495 nm (2.5 eV) luminescence of both crystalline and amorphous silicon dioxide samples from different origins. The following questions are raised: (a) is the band at 2.5 eV in spectrum of quartz related to the bulk or surface-related emission centers; (b) what is the annealing behavior of the 2.5 eV band in different $SiO_2$ samples; (c) what is the effect of crystalline structure on the 2.5 eV emission; (d) is there the 2.5 eV band in spectrum of amorphous silicon dioxide; (e) are the properties of this band similar to crystalline one; (f) is the band at 2.5 eV in luminescence spectrum of crystalline α-quartz is polarized at RT; (g) may the emission at 2.5 eV of silicon dioxide be related to $Li^+$ impurity.

## 2. Samples

Single crystal of α-quartz was grown using the hydrothermal method. Size of single crystal was approximately 1x1x3 cm (towards X, Y and Z axes). Part of this single crystal was milled in agate mortar into the powder with the characteristic particle sizes of 10-100 μm. Amorphous pure dry synthetic silicon dioxide powder had the particle sizes of 1-10 μm and possessed the weight percent of silicon dioxide of at least of 98 wt% (Russian state standard GOST 9428-73). The percent of volatile impurities was of 1.5 wt%. These impurities can be removed (at least partially) via the annealing in certain environment. The weight percent of silicon dioxide in the isotopically pure powder was of 99.984 wt%. The atomic percent of $^{28}Si$ isotope was of 99.91 at%. The total content of admixtures in powder was of 0.016 wt%. The annealing of all samples was carried out on the ambient air at 1000 ºC during the 6 hours.

## 3. Experimental setup

In present study the PCL spectra measurements of both crystalline and amorphous silicon dioxide samples were carried out under the action of short (~2 ns) high-current (~60 A) electron beam of RADAN-EXPERT accelerator [19,20] with particle energies of ~150 keV. Present excitation method with the spatial energy density of ~10 $MW/cm^2$ leads to the similar to [8] (up to $10^{18}$ $cm^3$) excitation density achievement. High excitation density is achieved, because the penetration depth of electrons with energies of ~150 keV is smaller in comparison with the penetration depth of electrons with energies of ~2 MeV [8]. Thus, although the peak current and total energy of electron beam in the present study are significantly lower, than in [8], the bulk energy and excitation densities are comparable. At the same time, the beam-induced damage in present study is decreased.

The cross-section area of electron beam was ~2-3 cm$^2$. Repetition rate of accelerator pulses was equal to 1 Hz. Maximum range of electrons with energies of 120-150 keV in silicon dioxide is of 30-60 μkm (the effective atom number $Z_{eff}$ of this material is equal to 10.8 [21]). Thus, the PCL is irradiated mainly from the thin near-surface layer of the sample. The luminescence from sample was directed into the entrance slit of an OCEAN USB2000 spectrometer via the quartz optical fiber. The fiber was used to reduce the effect of the electromagnetic interferences from accelerator on the spectrometer. The resolution of spectrometer was of 1.8 nm, and its spectral range was 200-800 nm. All spectra were corrected in accordance with the spectral sensitivity of the optical system. The CCD-line of spectrometer accumulated the light signal during the 30-sec exposure. Thus, the PCL spectra of each sample are the result of 29 luminescence pulses intensity summation. Thus, the contributions to the emission spectra during and immediately after the irradiation are also measured. The polarization of PCL was measured using the film polarizer with the optical transmittance (OT) in the region of 450-700 nm. All spectra were corrected in accordance with the OT of polarizer.

In the space between the electron tube of the accelerator and the sample under investigation the emission of second positive system of bands of molecular nitrogen N$_2$ (SPS) and first negative system of bands of molecular ion N$_2^+$ (FNS) occurs. This space is filled with ambient air and has a length of ~3 cm. The emission spectra during the excitation of air by the electron beam in the absence of a sample are shown in Fig 1 (a). Peaks of intensity at 316 nm, 337 nm, 358 nm, 367 nm, 376 nm, 380 nm, 391 nm, 400 nm and 406 nm correspond to electron-vibrational bands with unresolved rotational structure of C$^3\Pi_u$-B$^3\Pi_g$ transition (SPS) of molecular N$_2$. The peak at 426 nm corresponds to the (0.1) electron-vibrational band of B$^2\Sigma_u^+$-X$^2\Sigma_g$ transition of molecular ion N$_2^+$ [22].

The total energy of electron beam of RADAN-EXPERT accelerator may be strongly (up to 10 times) varied from pulse to pulse [19]. The emission spectra of SPS bands at different electron beam total energies were measured. On the basis of this result, the intensity ratios between 316, 337 and 358 nm SPS bands at different exciting electron beam total energies were obtained. The total energy of the electron beam was not controlled directly. The result is presented at Fig. 1 (b). The horizontal axis represents the intensity of 337 nm SPS band, and the vertical axes represent the corresponding intensities of 316 and 358 nm SPS bands at the appropriate total energy of electron beam. It is seen, that the of the 316 and 358 nm SPS band intensities are linearly proportional to the 337 nm SPS band intensity in the vicinity of experimental error. Consequently, the intensity ratio of these SPS bands is kept constant at the corresponding electron beam total energies range.

The primary mechanism of SPS $N_2$ emission excitation in ambient air conditions is the direct electron impact [23]. The variation of applied voltage to the emitting diode of the RADAN-EXPERT accelerator (~10 kV) is small in comparison to its absolute value (~150 kV) [19]. Thus, it can be expected, that the electron energy distribution function (EEDF) of the electron beam is not essentially varied from pulse to pulse. Consequently, it can be assumed in first approximation, that the number of electrons is varied only, not the EEDF. Accordingly, the intensities of SPS bands are proportional in first-order approximation to incident beam electrons number (fluence). Consequently, the intensities of SPS bands may be used for the electron beam fluence estimation in all the PCL spectra measurements.

The presence of SPS bands in PCL spectra makes possible the proper comparison of the different spectra measured, despite the electron beam total energy variation from pulse to pulse. Due to the linear dependence of the 316, 337 and 358 nm SPS band intensities on each other, all of these bands may be used for the PCL spectra intensity correction in accordance with the incident beam total energy.

### 4. Comparison of PCL spectra of all samples studied

Comparison of pure α-quartz single crystal PCL spectra before annealing (a), PCL spectra of the same crystal after annealing (b), α-quartz crystalline powder PCL spectra (milled part of the same non-annealed α-quartz single crystal) (c), PCL spectra of isotopically-pure $SiO_2$ powder with low concentration of impurities (0.016 wt%) after annealing (d), a-$SiO_2$ dry amorphous powder PCL spectra (98 wt% purity) before annealing (e) and the same amorphous powder PCL spectra after annealing (f) are presented in Fig. 2. The emission of isotopically-pure powder before annealing could not been detected due to its low intensity.

It can be seen from Fig 2, that the band with maximum intensity in the region of 495 nm (2.5 eV) exists in PCL spectra of all materials studied. In addition, all materials possess the band at 520 nm (2.37 eV). The band at shorter wavelength also exists in PCL spectra of all materials. It peaked at 415 nm (2.99 eV) in α-quartz single crystal PCL spectra before (Fig 2, (a)) and after annealing (Fig 2, (b)) and in the α-quartz crystalline powder PCL spectrum (Fig 2, (c)). PCL spectra of isotopically pure powder after annealing (Fig 2, (d)) and PCL spectra of amorphous powder before annealing (Fig 2, (e)) possess a band with maximum of intensity at 450 nm (2.8 eV). Also, the amorphous powder before the annealing possess a band at 650 nm (1.9 eV). This band is well-studied [4]. It is the sign of amorphous state. This band is related to non-bridging oxygen hole center (NBOHC) recombination [4]. This band is eliminated after annealing (Fig. 2 (f)). Also, after annealing the emission with constant intensity arises in the region of 250-450 nm in spectrum of amorphous powder (Fig. 2 (f)). The decrease of intensity in the region of 250-300 nm we attribute to the light absorption in the quartz fiber. Amorphous powder may be crystallized after annealing. In present work the structure of annealed amorphous sample did not been studied.

Ratios of the band intensities at 495 nm (2.5 eV) in PCL spectra of all materials studied ($I_{495\ nm}$) with respect to the maximal intensities at shorter wavelengths ($I_{shorter\ wavelength}$) in the same spectra are presented in Table 1. The following conclusions can be drawn. Firstly, let us consider the PCL spectra of α-quartz single crystal and α-quartz powder (milled single crystal). The intensity ratio of the 495 nm (2.5 eV) band with respect to intensity of 415 nm in PCL spectra of α-quartz single crystal ($I_{495\ nm}/I_{415\ nm}= 0.8\pm0.05$) is equal within the experimental error to the ratio between the intensities of the same bands in PCL spectra of the α-quartz crystalline powder ($I_{495\ nm}/I_{415\ nm} = 0.8\pm0.1$). It also can be seen, that the shapes of PCL spectra of these materials (Fig 2 (a), (c)) are still identical to each other. At the same time, the surface area of powder particles is of $10^4$-$10^6$ times higher, then the surface area of a single crystal, but the PCL intensity of crystalline powder is ~5 times lower than the single crystal PCL intensity. PCL intensity decrease in comparison with the crystalline sample is clearly explained, because the crystalline powder is opaque due to the strong light scattering.

Based on these facts one can be concluded, that the band at 495 nm (2.5 eV) in PCL spectra of α-quartz is not related to the surface emission centers. This band is related strictly to the bulk emission centers. The same conclusion can be drawn regarding the 415 nm (2.99 eV) PCL band.

Secondly, let us consider the case of amorphous powder and α-quartz single crystal before and after annealing. The intensity ratio $I_{495\ nm}/I_{415\ nm}$ is slightly changed for annealed α-quartz crystal and becomes equal to 1.0±0.1. A small increase (~20 %) of relative intensity of the 495 nm (2.5 eV) band with respect to other bands in PCL spectrum of quartz crystal is detected (Fig 2 (b)). The relative intensities of amorphous powder before ($I_{495\ nm}/I_{430\ nm}$=1.6±0.2) and after annealing ($I_{495\ nm}/I_{430\ nm}$=1.5±0.2) are equal within the experimental error. The present annealing behaviors of PCL spectra of crystalline and amorphous samples are very close to each other. This fact may be the manifestation of the same nature of the 495 nm (2.5 eV) bands in amorphous and crystalline $SiO_2$.

It is well-known, that the temperature of 573 ºC corresponds to the α-β polymorph transition in quartz structure at atmospheric pressure. The atmospheric heat of the β-quartz to temperature of 867 ºC leads to the transition of β-quartz to β-tridymite [24]. Accordingly, the α-quartz, annealed on air at 1000 ºC during the 6 hours, is the polymorphous crystal, composed of α- and β-quartz, and, probably, the β-tridymite. The polymorphism of the sample was manifested, particularly, in the high cracks concentration. This also led to the lower intensity of PCL of this sample. Unfortunately, we cannot exactly establish the distribution of polymorphs in this sample. At the same time, the emission at 650 nm did not been detected in crystal after annealing (Fig 2, (b)). Thus, the amorphous phase in this sample was negligible. Strictly speaking, it is wrong to name the annealed quartz crystal as "quartz". In present work "quartz" for this material is the conditional name.

It can be seen, that the annealed quartz crystal is sufficiently disordered in comparison with the non-annealed single crystal. The Fig 2 and Table 1 show, that the intensity 495 nm (2.5 eV) is the maximal intensity in PCL spectra of the disordered materials (powders and annealed polymorph crystal).

## 5. RT PCL spectra of α-quartz single crystal

Present section explores in detail the PCL spectra of α-quartz single crystal. The spectral instrument function (IF) of detection system was determined using the 473 nm solid-state laser. It was found, that the IF has the form of Gaussian function with the FWHM (Full Width on the Half of Maximum) of 3.8 nm. At the same time, the CCD-line of OCEAN USB2000 spectrometer possess a wavelength step of 0.37-0.38 nm. The contribution of CCD-line noise into the measured signal is very high at the long exposure times (up to 30 s). This leads to arising of the non-physical artificial distortions of the spectra measured. The characteristic scales of these distortions are of 0.37–1.1 nm. In order to improve the S/N ratio, filtration of PCL spectra was carried out. The filtration is based on numerical convolution of the spectral data from spectrometer with its own IF. The idea of present filtration is to remove the artificial oscillations with periods of 0.37-1.1 nm from spectra. It may be imagined, as the additional device in scheme, with IF, which is identical to the IF of the OCEAN USB2000 spectrometer.

The Fourier images of Gauss functions are still the Gauss functions. In the Fourier space, the convolution is the algebraic product of Fourier images of the functions (initial signal and IF's) [25]. Thus, the Fourier transform of final IF is the product of two equal Gauss functions, and the final IF is still the Gaussian function with FWHM of $\sqrt{2}$ times wider, than the IF of spectrometer. Gauss form of IF is very convenient, because the artificial oscillations are not arising in the signal after convolution [26]. It is not true, for example, for the rectangular IF, because the convolution of signal with function of this type produces the side lobes [26]. The FWHM of final IF after the processing is of 5.4 nm. At the same time, the S/N ratio is much higher (~20 times), than in the initial data.

The spectra of α-quartz single crystal in the region of 475-500 nm are presented in Fig 3 (a). Solid curve shows the average of 7 spectra measured. All spectra were filtered according to the method discussed above. Error is estimated as a root-mean square deviation between the average spectrum and spectra measured. In the inset (Fig. 3 (b)) the spectra of 473 nm laser (IF of spectrometer) (1), its Gauss fit (2) and the final IF of the measurement system after filtering (3), are presented. The measurement system with filtering may be considered as the system with one additional "filtering" device with Gauss IF. It is seen, that final IF has the Gauss form with FWHM of 5.4 nm.

The low statistics (7 measurements only) was used (a) to prevent the radiation-induced damage in quartz [27] and (b) due to the low quantity of the accelerator pulses with higher energies of electron beam, which are lead to the high-intensity PCL with appropriate S/N level. It is seen (Fig 3, (a)), that the 495 nm (2.5 eV) PCL band possess a structure, containing the three separate peaks. The peaks are positioned at 480±2 nm (2.58±0.01 eV), 487±2 nm (2.55±0.01 eV) and 493±2 nm (2.52±0.01 eV). After the processing, the intensity error does not exceed the 2%. Peaks are equidistant: energy difference $\Delta E$=0.03±0.01 eV=240±80 cm$^{-1}$. This quantity is close to the lithium oxide molecular vibration frequency. For example, one is equal to the Li$_2$O vibration frequency - 230 cm$^{-1}$ (bend mode) [28]. Also, this quantity lies in the region of optical phonon modes $E$(TO) (265 cm$^{-1}$) and $E$(LO) (265 cm$^{-1}$) of α-quartz [29].

The RT spectra of 7 times averaged α-quartz single crystal in the region of 300-600 nm are presented in Fig 4 (a). Background fill (orange) shows the estimation of root-mean square std. deviation. The bands at 316, 337, 358 and 380 nm are assigned to SPS and FNS emission of molecular nitrogen. There is a separate band at 495 nm (2.5 eV). To figure out, is this band of the same nature to the bands at 2.5 eV in [6-8], the polarization of the α-quartz single crystal emission of in region of 450-700 nm was measured.

PCL spectra of α-quartz single crystal in region of 450-700 nm in polarization parallel to the X-axes, measured from the XZ plane of crystal, are presented in Fig 4 (b), in polarization parallel to the X-axes, measured from the XY plane of crystal, are presented in Fig 4 (c) and in polarization parallel to the Y-axes, measured from the XY plane of crystal, are presented in Fig 4 (d). The intensity of emission of α-quartz single crystal in the polarization parallel to Z axes from XZ and YZ crystal planes was weaker than the noise level. Thus, one can be concluded, that the emission of α-quartz single crystal in the region of 495 nm (2.5 eV) is polarized in the XY crystalline plane. The noise level is of ~10%. Consequently, the polarization degree $(I_{\parallel}-I_{\perp})/(I_{\parallel}+I_{\perp})$ is not less than the 0.8. The $I_{\parallel}$ denotes here the PCL intensity with polarization parallel to XY plane, $I_{\perp}$ denotes the PCL intensity with polarization perpendicular to XY plane. This result is in accordance with the data on polarization of 2.5 eV emission of α-quartz from [7,8].

The coincidence of polarization properties of the low-temperature 2.5 eV emission band in quartz, observed in [7,8], and the 495 nm (2.5 eV) RT band in PCL spectrum of α-quartz single crystal in present study (Fig. 2 (a), Fig 3 (a), Fig. 4 (a)) may be the manifestation of the same origins of these emissions. We suppose, that the strict polarization of 495 nm (2.5 eV) PCL band in XY plane is an important argument in favor of this assertion. Further research will shed more light on this phenomenon.

The polarization studies of structure of the band at 495 nm (2.5 eV) in PCL spectra of α-quartz are presented at Fig. 5. It can be seen, that the structure is well-observed in polarization parallel to X axis. The blurring of the structure of the 495 nm (2.5 eV) band in the polarization parallel to Y axis may be related to the lower S/N ratio in this case.

## 6. Discussion

The present data is insufficient for the direct establishment of the nature of 495 nm (2.5 eV) silicon dioxide emission. At the same time, we try to propose the appropriate model of radiation process. In order to explain the results obtained we suggest the possible hypothesis which is related to the $Li^+$ ions participation in the radiation process. The verification of present hypothesis is the problem for further research.

Firstly, let us consider the emission at 415 nm (2.99 eV) in RT PCL spectra of α-quartz single crystal. The position of this band lies exactly in the region of the STE emission [1,3,4,6,8,10,11,13]. At the same time, it is well-known, that the STE in α-quartz is strongly quenched at RT [10]. Thus, the association of this band with the STE recombination is insufficiently justified. However, we try to explain the data obtained in accordance with the assumption, that the band at 415 nm (2.99 eV) is related to STE recombination.

Next, let us consider the structure of the 495 nm (2.5 eV) emission band in PCL spectra of α-quartz single crystal (Fig. 3,5). The separation between peaks in this structure is equal to the optical phonon mode of α-quartz and, at the same time, close to the lithium oxide molecular vibrations. The observed structure cannot be attributed to the Stark splitting of the defect or exciton energetic levels. Indeed, the internal defect-mediated crystalline electric fields $E_c$ in α-quartz are of the order of 10-150 kV/cm [30]. Consequently, the dipole moment $d \approx \Delta E/E_c$ of the corresponding quantum state is of 100-1500 D. At the same time, the characteristic dipole moments in α-quartz are of the order of 1-10 D [31]. Also, the structure of 495 nm (2.5 eV) band cannot be attributed to the $O_2$ molecule vibrations, because their frequency is of ~1600 cm$^{-1}$ [32].

The relation of the 495 nm (2.5 eV) band structure to the lithium oxide molecular vibrations is in agreement with the results of [15]. It is important, that the radiation processes in quartz at RT may be strongly different in comparison to low-temperature cases [6-8,10] due to the (a) strong STE emission quenching at RT [10] and (b) growth of probabilities of the impurity-related processes due to the increase of impurity ions mobility.

The probable model of the $SiO_2$ luminescence in the region of 495 nm (2.5 eV) at RT, which is based on the dominant role of $Li^+$ ions, is follows. Alkali ions exists in all $SiO_2$ samples as the charge compensators for the $Al^{3+}$ ions – the most abundant impurity in $SiO_2$. Interstitial $Li^+$ ions are located adjacent to substitutional $Al^{3+}$ ions in as-grown quartz [33]. If the sample temperature is near or above 200 K, the $Li^+$ ions can be moved away from these trapping sites after the high-energy irradiation [16]. Evidently, they are delocalized after irradiation at RT.

The irradiation in present study produces a high density of excitation (as well as in [8]). Consequently, the $Li^+$ ions should be efficiently delocalized from their traps. At the same time, the high-current electron beam irradiation leads to the formation of STEs [8]. According to [13], the STE in α-quartz is accompanied by the Si-O bond rapture and the transient formation of the NBO. The rapture of bond may be caused, for example, by the direct electron impact excitation of the Si-O bond into the excited antibonding state. The formed negatively charged oxygen ion $O^-$ ($NBO^-$-center) is the attractor both for the hole and $Li^+$ ion. If the hole is attracted to the $NBO^-$-center and electron is being localized at neighboring Si atom, the STE may be formed [13].

At the same time, if the $Li^+$ ion is attracted to the $NBO^-$-center, the $NBO^-$-$Li^+$ complex may arise. The presence of these complexes in α-quartz is noted in [34]. After recombination, these complexes, presumably, are destroyed. It is in consistence with the impossibility of the permanent NBOs existence in perfect α-quartz structure [4] due to steric limitations.

Thus, the proposed model is based on the assumption, that the formation of $NBO^-$-$Li^+$ complexes possess the preliminary stages exactly similar to the STE formation. Consequently, this model requires the strict direct proportionality between the emission intensity at 495 nm (2.5 eV) and the intensity of STE emission before the saturation of $Li^+$-related centers concentration.

Assuming, that the 415 nm band is related to STE, we found the linear proportionality of the 495 nm (2.5 eV) emission intensity to the intensity of STE recombination. The PCL spectra of the same quartz single crystal under the action of accelerator beams with different total energies are shown in Fig. 6 (a). Every spectrum was measured with exposure time of 30 s. It is seen, that the lowest and the highest intensities are related as ~1:10. Intensities of 490 and 415 nm bands of quartz single crystal and 315, 337 and 380 nm bands of SPS $N_2$ at different electron beam total energies are obtained from this data. The dependence of the intensity of 490 nm (2.53 eV) PCL band of quartz single crystal and 315, 337 and 380 nm SPS bands on the intensity of 415 nm (2.99 eV) band is shown at Fig. 6 (b). It is clear, that the intensity of 490 nm (2.53 eV) PCL band of quartz is linearly proportional to the intensity of 415 nm (2.99 eV) band in the certain range of beam total energies. Also, all the SPS bands are directly proportional to the intensity of 415 nm (2.99 eV) band. The last fact (direct proportionality of the 415 nm band intensity to the intensities of SPS bands) shows, that the intensity of band at 415 nm is proportional to the total energy of incident electron beam.

However, according to present model, the saturation of 495 nm (2.5 eV) emission band intensity should occur at certain excitation density in comparison to STE emission one. It was the fact that was established in [8].

The present model explains why is the emission at 495 nm (2.5 eV) the most intensive one in spectra of disordered materials (powders and annealed quartz crystal). Unlike the case of quartz, the NBOs can permanently presence in the amorphous structure [4]. The manifestation of this fact is the 650 nm band in PCL spectrum of non-annealed amorphous $SiO_2$ powder (Fig 2 (e)). This band is well-studied and attributed in the most works to NBOHC [4]. Thus, the transient Si-O bond rapture during the STE excitation is not the only mechanism of the NBOs in amorphous samples. Consequently, the preexisting NBOs in $a$-$SiO_2$ samples also take part in the $NBO^-$-$Li^+$ complexes formation.

According to this model, the annealing behavior is simply explained. The annealing at 1000 ºC have no significant effect on the $Li^+$ ions localization. At the same time, the annealing of crystalline sample leads to an increase of defect concentration and probable amorphisation. Thus, the increase of pre-existing NBOs concentration is expected in annealed quartz crystal in comparison with the non-annealed one.

At the same time, an increase of the pre-existing NBOs concentration in amorphous sample after annealing is negligible in comparison to the case of annealing of the single crystal. Accordingly, due to this fact, the intensity ratio between the 495 nm (2.5 eV) and 430 nm (2.88 eV) bands is became unchanged after annealing for amorphous sample. The common increase of intensity after annealing may be explained by the quenching impurity elimination (for example – water vapors). In [35] the close annealing behavior of the 2.5 eV band of quartz was established.

Let us consider the polarization of 2.5 eV emission band. It is well-known [36], that the quartz structure possesses the two types of channels ("*small*" and "*large*") parallel to *c* (Z) axis. Both two channels are strictly perpendicular to the XY plane. It is shown, that the interstitial $Li^+$ ion can travel through the "*large*" channel after irradiation at RT [37]. When the $Li^+$ ions meets the $NBO^-$ center, one is being positioned exactly at the shortest possible distance to the $O^-$ ion due to the Coulomb attraction. The shortest distance is realized when the $Li^+$ ion is positioned in the same XY plane with corresponding $O^-$ ion. It is clear, that the dipole moment of this complex lies exactly in XY plane. Thus, the recombination luminescence will be still polarized at XY plane. This fact is experimentally confirmed in [7,8] and in the present study.

The final question is concerned with the structure of 495 nm (2.5 eV) band (Fig. 3 (a), Fig. 5). One may be explained as the electron-vibration transition in $NBO^--Li^+$ complex. In favor of this hypothesis is the fact that the separation between peaks is close to the frequency of lithium oxide molecular vibrations. To our knowledge, the studies of $NBO^--Li^+$ complex vibration frequencies have not been carried out at the present moment. Consequently, the further research is required to verify this hypothesis. The probable scheme of radiative electron-vibrational transition from the excited $NBO^--Li^+$ configuration to the perfect lattice (without NBO, with the restored Si-O-Si bond) and interstitial $Li^+$ ion is shown in Fig. 7. The horizontal axis represents the Li-O distance and vertical axis represents the potential energy of system. The transition occurs from the ground (v=0) and first excited (v=1) vibrational state of $NBO^--Li^+$ complex. Three bands arise due to the Franc-Condon factors. The ground state is the repulsive one and have no minimum of potential energy.

The sensitivity of the 2.5 eV emission band in $Li^+$-excess quartz sample to the radiation dose may be explained as the radiolytic formation of the long-lived interstitial $Li_xO_y$ molecules in structure of quartz. One may lead to the change in structure of 495 nm (2.5 eV) emission band due to the increase of vibrating Li-O species number. This question requires the further studies.

It can be seen, that the $Li^+$-based model may explain all the present experimental data and the results of [7,8,15,35]. However, the present model contradicts to the results of [9-12]. The measurements of quartz luminescence in present study and in [15,35] are conducted at RT under the action of high-power irradiation. At the same time, the influence of the $Li^+$ ions on the behavior of 495 nm (2.5 eV) luminescence at cryogenic temperatures is become unclear. The further studies are required to resolve the present contradiction.

## 7. Conclusions

1. The RT PCL-spectra of α-quartz single crystal before and after annealing, milled α-quartz crystalline powder, isotopically-pure $SiO_2$ powder after annealing, pure dry amorphous $SiO_2$ powder before and after annealing possess a PCL band with maximum intensity at 495 nm (2.5 eV). The band at 495 nm (2.5 eV) is the most intensive one in PCL spectra of disordered materials (non-crystalline powders and annealed quartz crystal).

2. The intensity ratios of the bands at 415 and 490 nm in PCL spectra of α-quartz single crystal and α-quartz crystalline powder (milled α-quartz single crystal) are equal within the experimental error. This is the direct evidence of the bulk nature of corresponding emission centers.

3. The annealing behaviors of the 495 nm (2.5 eV) PCL band intensities of crystalline and amorphous $SiO_2$ samples are very close to each other. This fact may be the manifestation of the identical nature of these bands in amorphous and crystalline $SiO_2$.

4. The structure of the 2.5 eV PCL band in spectra of α-quartz single crystal is detected. It consists of three peaks: at 480±2 nm (2.58±0.01 eV), 487±2 nm (2.55±0.01 eV) and 493±2 nm (2.52±0.01 eV). Energy separation between the peaks lies in the region of lithium oxide molecular vibrations energies and the energies of optical phonon modes of α-quartz.

5. The polarization of the 495 nm (2.5 eV) emission band in XY crystalline plane in RT PCL spectra of α-quartz single crystal is detected. Structure of the 495 nm (2.5 eV) is well-observed in polarization parallel to X plane and blurred in polarization parallel to Y plane.

6. The linear proportionality of the intensity of 490 nm (2.53 eV) emission band to the intensity of 415 nm (2.99 eV) band is detected in certain total electron beam energy range.

7. The explanation of the results obtained is given based on assumption that the origin of 495 nm (2.5 eV) emission in $SiO_2$ is related to non-bridging $O^--Li^+$ complexes recombination.

## 8. Acknowledgement

The present work is supported by the Russian Science Foundation (project No. 14-22-00273).

**Table 1**.

Ratios of the 495 nm (2.5 eV) band intensities of silicon dioxide samples $I_{495\ nm}$ with respect to maximal intensities $I_{shorter\ wavelength}$ in PCL spectra of the same materials at shorter wavelengths

| Material | Fig. 2 | Shorter wavelength | $I_{495\ nm}/I_{shorter\ wavelength}$ |
|---|---|---|---|
| α-quartz single crystal | (a) | 415 nm (2.99 eV) | 0.8±0.05* |
| annealed α-quartz single crystal | (b) | 415 nm (2.99 eV) | 1.0±0.1 |
| α-quartz crystalline powder | (c) | 415 nm (2.99 eV) | 0.8±0.1 |
| annealed isotopically-pure $SiO_2$ powder (total content of admixtures of 0.016 wt%) | (d) | 430 nm (2.88 eV) | 1.2±0.2 |
| amorphous $SiO_2$ powder (98 wt% purity) | (e) | 430 nm (2.88 eV) | 1.6±0.2 |
| annealed amorphous $SiO_2$ powder (98 wt% purity) | (f) | 430 nm (2.88 eV) | 1.5±0.2 |

*The accuracy of the $I_{495\ nm}/I_{415\ nm}$ ratio determination in case of α-quartz single crystal is limited by the structure of the 495 nm emission band (see Sec. 5). The accuracy of present ratio is low enough to neglect the structure. Real error level of our α-quartz single crystal PCL spectra measurements is several times better.

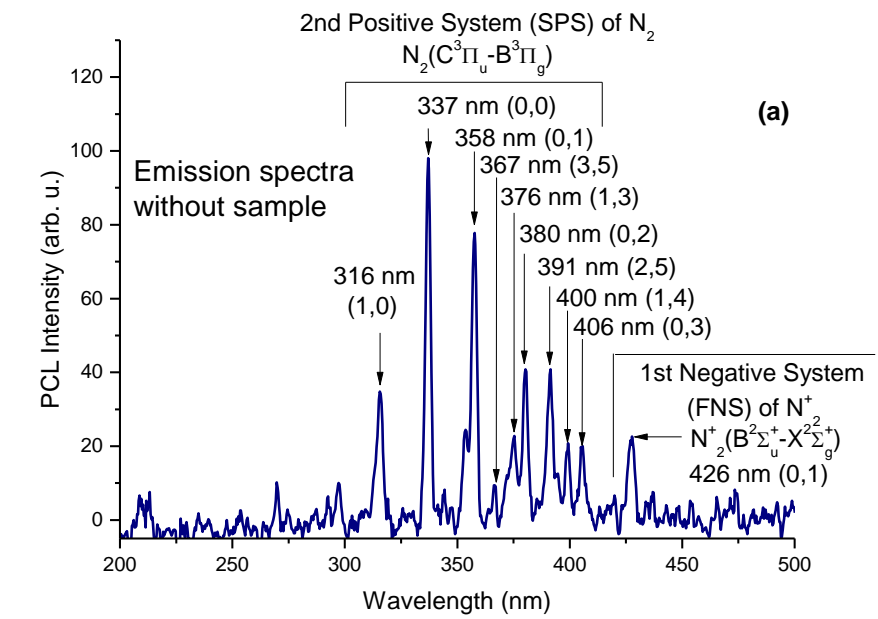

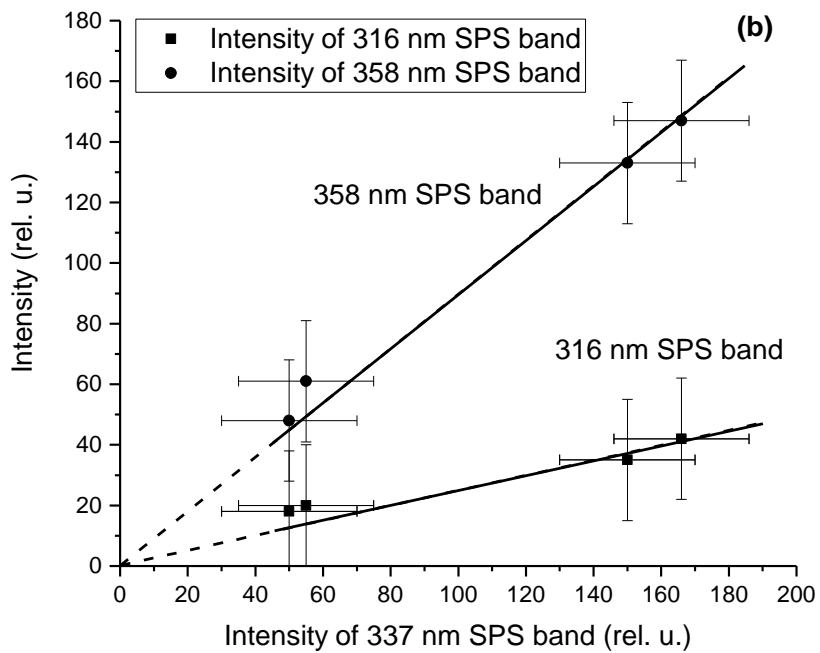

**Fig. 1**. (a) Emission spectrum in the absence of sample in setup. Spectrum is dominated by the SPS and FNS bands of $N_2$; (b) Intensities of 316 and 358 nm emission bands of SPS $N_2$ with respect to the 337 nm SPS band intensity at different total energies of incident electron beam.

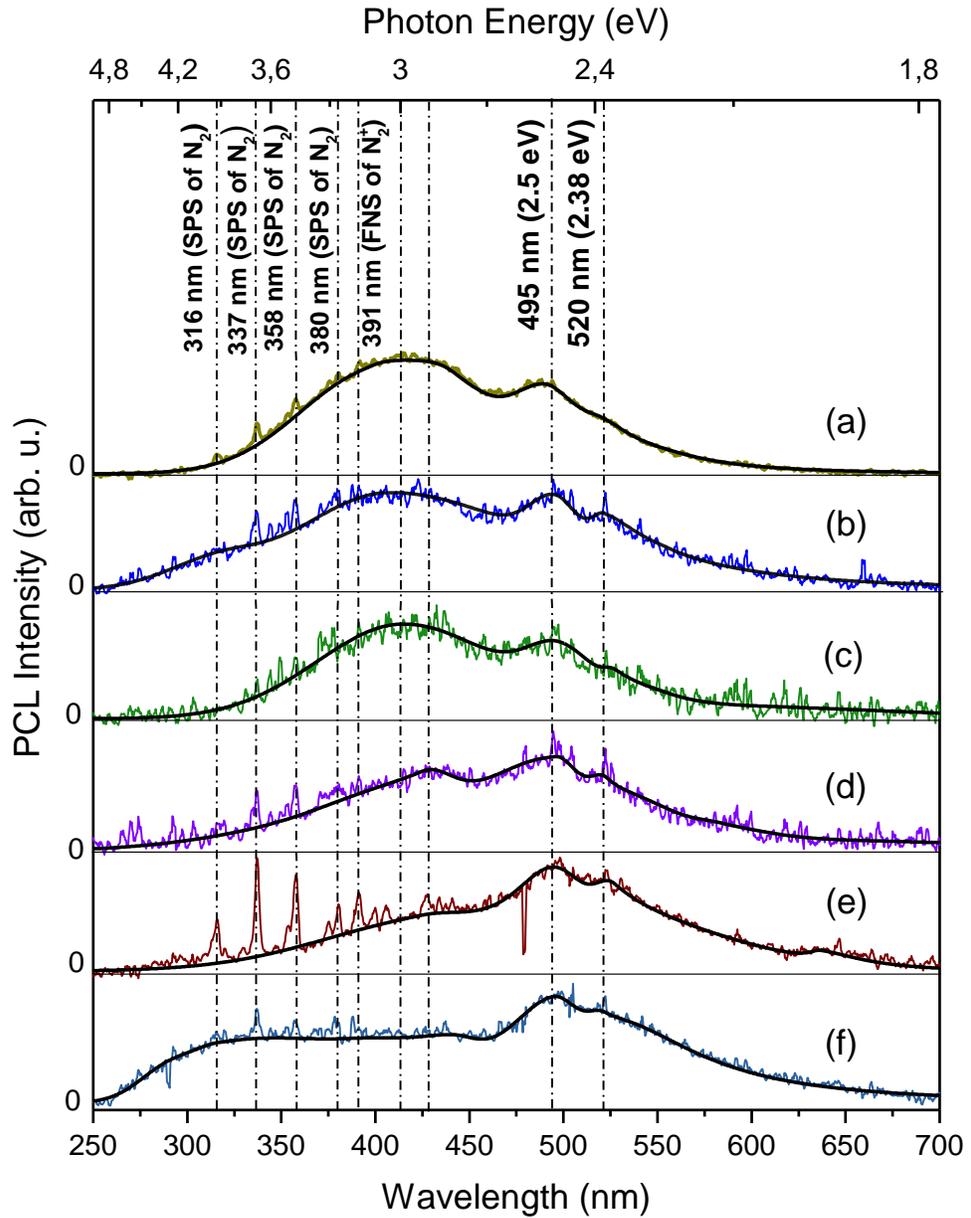

**Fig. 2.** Room-temperature PCL spectra of α-quartz single crystal before annealing (a), α-quartz single crystal after annealing (b), α-quartz crystalline powder (c), isotopically-pure SiO₂ powder with purity of 99.984 wt% after annealing (d), amorphous SiO₂ powder with purity of 98 wt% before (e) and after annealing (f). Bands with maxima at 316, 337, 358, 367, 381, 400 and 425 nm are related to molecular nitrogen SPS and FNS emission [22].

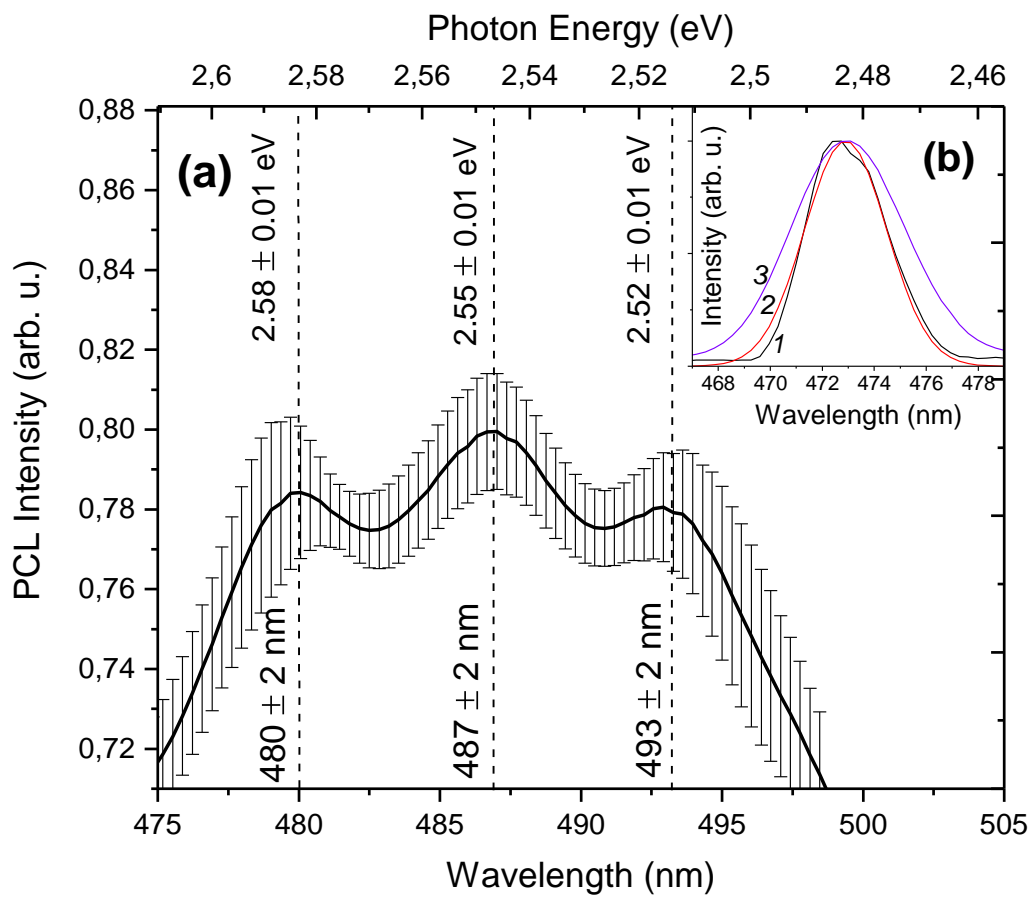

**Fig. 3.** (a) Room-temperature PCL spectrum of α-quartz single crystal in the region of 475-598 nm. Solid curve shows the average spectrum. Averaging is carried out on the smoothed spectra (explanation in text). Error bars shows the estimation of std. deviation. (b) (inset) Spectrum of 473 nm solid-state laser. 1 – laser spectrum measured (IF of spectrometer), 2 – Gaussian fit of this spectrum with FWHM of 3.8 nm, 3 – Resulting IF after the convolution of spectrum (1) with Gaussian function (2). The final FWHM is of 5.4 nm.

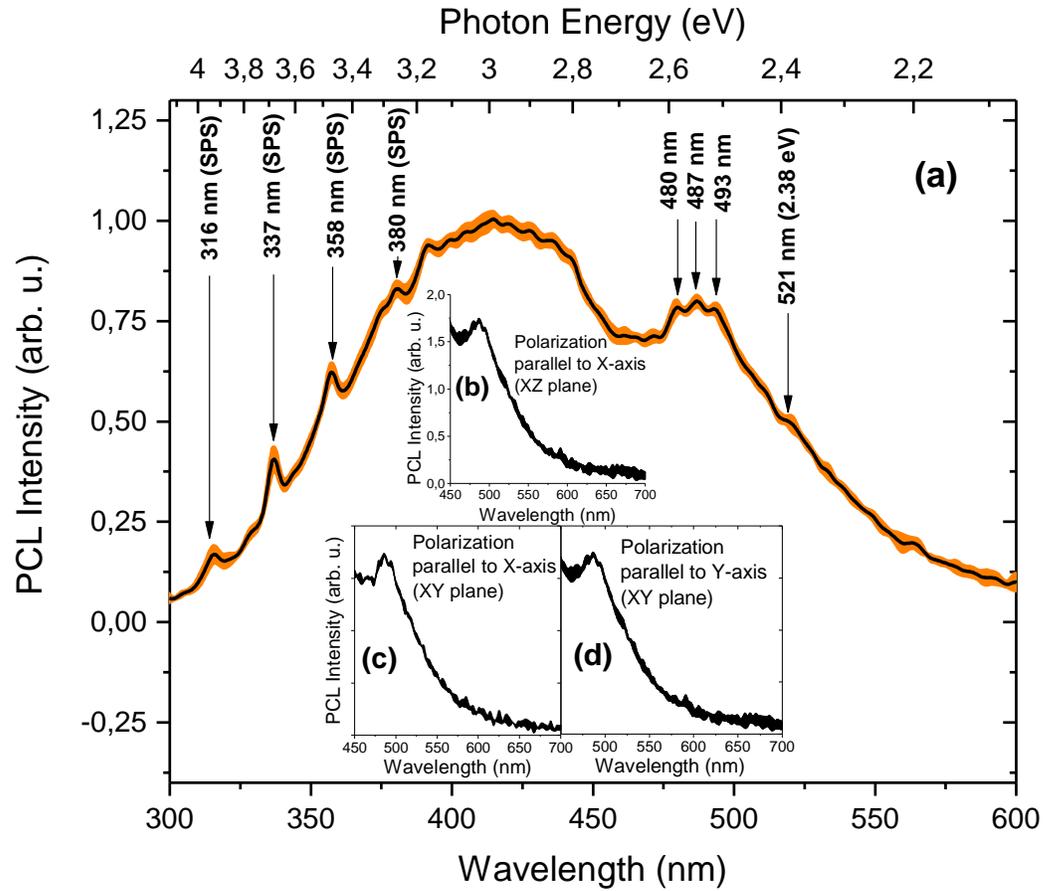

**Fig. 4**. (a) Room-temperature PCL spectra of α-quartz single crystal. Solid (black) curve – average spectrum, Background (Orange) – root-mean square deviation estimation. Insets: PCL spectra of α-quartz single crystal in the region of 450-700 nm in polarization parallel to X crystal axis (emission measured from XZ plane) (b), parallel to X crystal axis (emission measured from XY plane) (c) and parallel to Y crystal axis (emission measured from XY plane) (d).

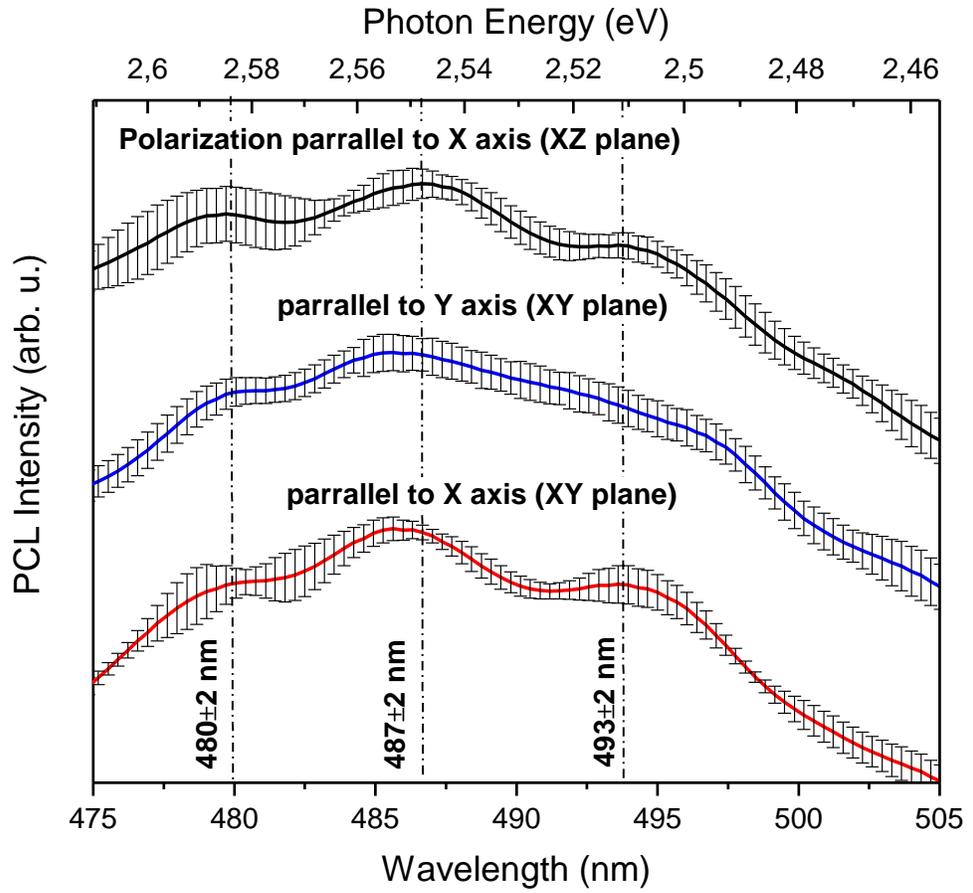

**Fig. 5.** Structure of the 495 nm (2.5 eV) PCL band of α-quartz in different polarizations.

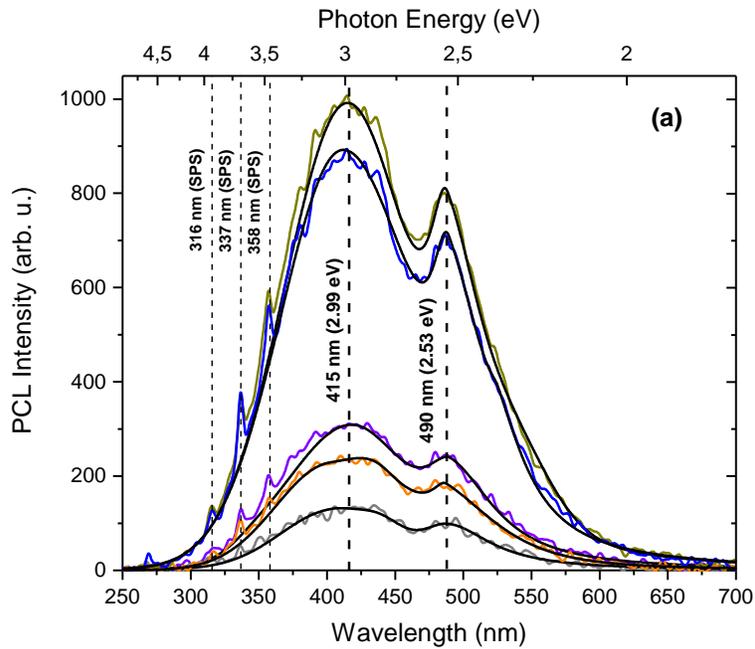

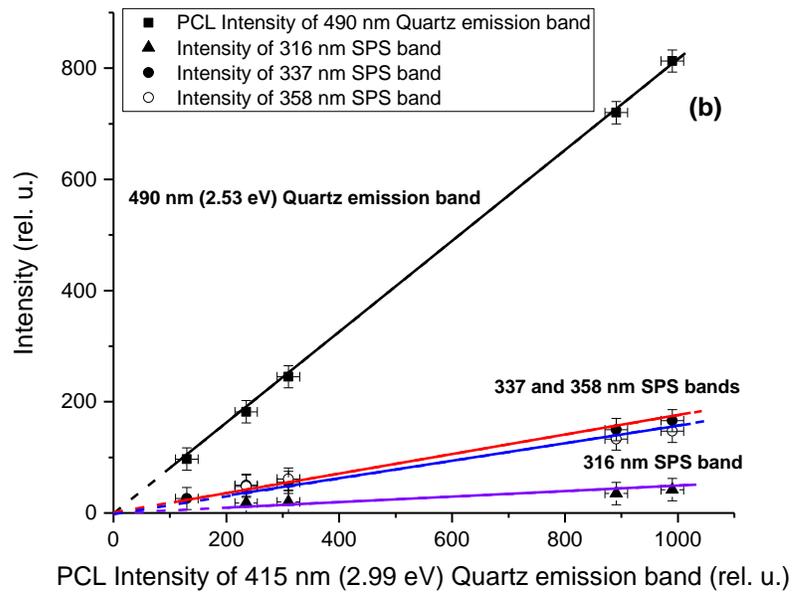

**Fig. 6.** PCL spectra of α-quartz single crystal under the action of electron beams with different total energies (a); The intensities of 490 nm (2.53 eV) and 316, 337 and 358 nm SPS N$_2$ emission bands with respect to the intensity of the 415 nm (2.99 eV) α-quartz emission band (b). These dependences are plotted on the basis of spectra from fig. 6 (a).

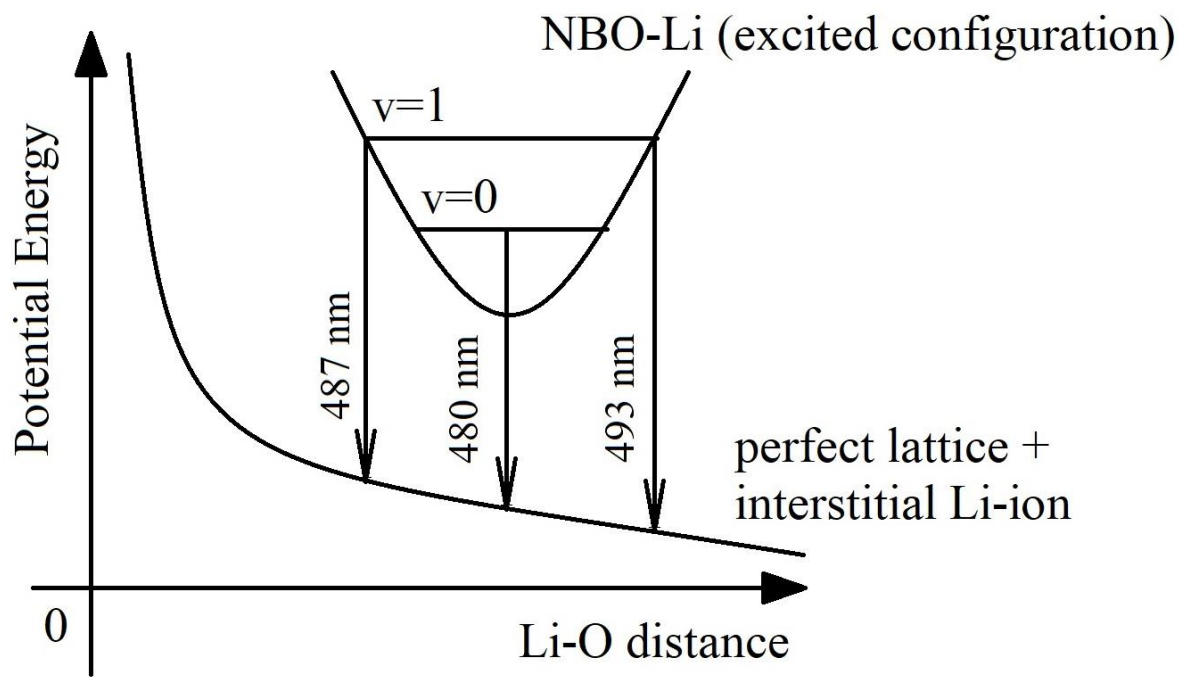

**Fig. 7**. Probable scheme of the NBO⁻-Li⁺ complex radiative recombination with structure of 2.5 eV (495 nm) emission band formation. In the upper curve "v" represents the vibrational quantum number.